\documentstyle[12pt]{article}
\renewcommand{\thesection}{\arabic{section}}
\renewcommand{\theequation}{\arabic{equation}}
\renewcommand{\theequation}{\thesection-\arabic{equation}}
\bibliography{plain}

\title{\bf  Restoration of Macroscopic Isotropy on $(d+1)$-Simplex Fractal Conductor
Networks   }
\vspace{20mm}
\author{M. A. Jafarizadeh$^{a,b,c}$ \thanks{E-mail: jafarzadeh@ark.tabrizu.ac.ir}\\
\\
\\
$^a${\small Department of Theoretical Physics and Astrophysics, Tabriz University, Tabriz 51664, Iran.} \\   
$^b${\small Institute for Studies in Theoretical Physics and Mathematics, Tehran 19395-1795, Iran.} \\
$^c${\small Pure and Applied Science Research Center, Tabriz 51664, Iran.}}

\begin{document}
\maketitle
\vspace{15mm}
\begin{abstract}

Restoration of macroscopic isotropy has been  investigated in $\bf (d+1)$-simplex 
fractal conductor networks via exact real space renormalization group 
transformations. Using some theorems of fixed point theory, it has been shown
very rigoroursly that the macroscopic conductivity becomes isotropic for large
scales and anisotropy vanishes with a scaling exponent which is computed
exactly for arbitrary values of $\bf d$ and decimation numbers $\bf b=2,3,4$ and
$\bf 5$.


{\bf PACs Index: 64.60.AK and 05.50}

\end{abstract}

\vspace{70mm}

\section{INTRODUCTION}

Restoration of isotropy in an anisotropic system is of great interest in a 
variety of disciplines where much attention has been focused on it, particularly
on the problem of diffusion in inhomogeneous  materials \cite {Smith,Lobb,Haus}.
In general, diffusion on lattices can be formulated in terms of an {\bf AC} electric
problem and {\bf DC} electric response in a percolating structures can be viewed 
as a very special case of diffusion in disordered medium \cite{Haus,Cle,Derri}.
The purpose of this paper is to investigate the restoration of macroscopic 
isotropy in $\bf (d+1)$-simplex fractal conductor networks with microscopic
anisotropy. In general, deterministic fractal lattices\cite{Mandel,Sch,Gefen,Jul}, as proposed by
Kirkpatrick, mimic some properties of percolation clusters in random media and 
disordered systems\cite{Gefen}, and among  fractal objects, the $\bf (d+1)$-simplex fractal is the   
simplest one to study various physical problems from random walk 
\cite{Jaf,Jafa,Jafar,Jafari} to electrical problem on it\cite{Ste,Pak}.  
Using the exact renormalization technique based on
the minimization of total dissipative power ({\bf  TDP}) in these networks, we present 
a rigorous proof that the conductivity becomes isotropic for large scales, 
and anisotropy vanishes with a scaling exponent $ \bar{\lambda}$,
as $\bf L^{-\bar{\lambda}}$\cite{Vanni}. We exactly compute $\bar{\lambda}$
 for arbitrary values of $\bf d$ and decimation numbers $\bf b=2,3,4$ and $\bf 5$.
The contents of this paper is as follows:

Section II presents a brief description of $\bf (d+1)$-simplex fractals with
decimation number $\bf b$ together with an explanation of labelling 
their subfractal and vertices with the partitions of positive integers\cite{Andr,Ham}, where
this coding plays a very important role throughout the article.  
In section III we consider the most general network that can be built in a
deterministic way, by putting circuit elements on the bonds of $\bf (d+1)$-simplex
fractal of a given generation $\bf n$ with decimation number $\bf b$. In order for the
self-similarity of the structure to be preserved in the presence of 
anisotropy at microscopic level, the nature of the circuit elements, namely 
its resistances, must depend on the orientation of the bonds. It is clear
that in $\bf (d+1)$-simplex there are $\bf \frac{d(d+1)}{2}$ different orientations. 
Then we try to establish recursion equations for the connection resistances which 
represent the conductivity of these networks, on two successive length scales
$\bf L$ and $\bf L^{\prime}=bL$. In general, these recursion relations are very
involved. Fortunately, we do not need to have the explicit form of these recurrence
equations, for the investigation of the restoration of isotropy. All we need
here is the general properties of these maps, which can be
obtained through some physical requirements and assumptions. It should be
stressed that these circuits are not fictitious, since $\bf (d+1)$-simplex
fractals are embedible in Euclidean $\bf 2$-dimensions, hence they can be
considered as two-dimensional networks, see Fig. {\bf 1}.
Section IV is devoted to a very rigorous proof of the uniqueness of the
fixed point of the real space renormalization group transformation of the
ratios of the connection resistances. Here in this section we show that all
flows of the real space renormalization group transformation of the connection
resistances, stemming from the finite physical region of connection resistance
space, diverges to a direction which makes equal angle with all 
coordinates axes. The proof is based on some theorems and definitions of fixed
point theory of the maps on complete metric spaces with the Hilbert metric. 
We have quoted the required theorems without presenting their proofs, since
this section would be otherwise more mathematical in style. We refer the readers
to reference \cite{Vas}
for proofs of all theorems and for more details. Those readers who are only interested
in the results of this section can skip it.
In section V by minimizing the {\bf TDP} in isotropic state, we get linear equations
for the inner inward flowing currents in terms of input currents with Lagrange
multipliers as their coefficients. Then using  $\bf S_{(d+1)}$ symmetry group of
the $\bf (d+1)$-simplex, we suggest an ansatz for the Lagrange multipliers
which leads to determination of the inner flowing currents in terms of the
input one for any values of $\bf d$ and decimation number $\bf b=2,3,4$ and $\bf 5$.
Section VI contains the main results of the article. Here in this section, by
linearising the recurrence relation of the connection resistances near the isotropy state,
we calculate power scaling exponent and the scaling exponent of the suppression of the
anisotropy, for arbitrary values of $\bf d$ and decimation numbers $\bf b=2,3,4$ and
$\bf b=5$, which are in agreement with the results of references \cite{Cle,Vanni,Barl,watan,hatto,hattor}
in special cases. Also these results hold true  for $\bf (d+1)$-honeycomb 
fractal conductor network with decimation number $\bf b=2 $, see Fig. {\bf 2}, which 
is in agreemet with reference \cite{Pak} for $\bf d=2$ and $\bf b=2$ case. 
The paper ends with a brief conclusion.

\vspace{10mm}
\section{ (d+1)-Simplex Fractals}
\setcounter{equation}{0}

$\bf (d+1) $-simplex fractal is a generalization of a two dimensional
Sierpinski gasket to $\bf d $-dimensions such that its subfractals are 
$\bf (d+1)$-simplices or $\bf d$-dimensional polyhedra with
$\bf  S_{(d+1)} $-symmetry. In order to obtain a 
fractal with decimation number $\bf  b $, we choose a $\bf (d +1)$-simplex
and divide all the links (that is the lines connecting sites ) into $\bf  b $
parts and then draw all possible $\bf d$-dimensional hyperplanes through the
links parallel to the transverse $\bf  d $-simplices. Next, having omitted
every other innerpolyhedra, we repeat this process for the remaining simplices
or for the subfractals of next higher generation. This way through
$\bf (d+1)$-simplex fractals are constructed. In order to calculate the fractal
dimension, also to determine the current distribution, it is convenient to
label subfractals of generation ($n$+1) in terms of partition of $\bf (b-1)$ into
 $\bf (d+1)$ positive integers $\bf  \lambda_1,\lambda_2,...,\lambda_{d+1}$.
 Each partition represents a subfractal of generation $\bf n$, and $\bf \lambda$ shows
 the distance of the corresponding subfractal from $\bf d$-dimensional
 hyper-planes which construct the $\bf (d+1)$ simplex. On the other hand,
 each vertex denoted by partition of $\bf b$ into $\bf (d+1)$ non-negative
integers $\bf \eta_1,\eta_2,\cdots,\eta_{d+1}$ and obviously the $\bf i$-th vertex
of subfractal $\bf (\lambda_1,\lambda_2,\cdots,\lambda_{d+1})$ is denoted by
$\bf \eta_j= \lambda_j+\delta_{i,j}$, where $\bf j=1,2,\cdots, d+1$.
As an illustrating example we show in Fig. {\bf  3} the method of labelling
a Sierpenski gasket with decimation number $\bf b=3$.

Obviously the number of all possible partitions is equal to the distribution
of $\bf (b-1)$ objects amongst $\bf (d+1)$ boxes, which is the same as the Bose-Einstein 
distribution of $\bf (b-1)$ identical bosons in $\bf (d+1)$ quantum states. This is 
equal to

\begin{equation}
C=\frac{(b+d-1)!}{(b-1)!.d!}.        
\end{equation}
    
As is well known, the fractal dimension $\bf D_f$ of a self similar object          
is defined according to \cite{Mandel}
$$
NL^D_f=1
$$
where $\bf N$ is the number of similar objects, up to translation and rotation, 
here being equal to the number of subfractals of generation $\bf n$, and
$\bf L $ is the scale of subfractal of generation $\bf n$. 

Hence 
 
$$ 
 N=C^r     ,  L=b^-r
$$
Therefore, 

$$
D_f=\frac{\ln c}{\ln b},
$$ 
or
\begin {equation}
D_f=\frac{\ln(\frac{(b+d-1)!}{(b-1)!)}}{\ln b}.
\end {equation}

\section{Fractal Connection Resistances and their Exact \newline
Renormalization Group Transformations}  

\setcounter{equation}{0}

A two-dimensional anisotropic $\bf (d+1)$-simplex resistor network consists
 of $\bf (d+1)$ nodes, with $\bf I_i$ denoting the amount of current injected 
into the network through the node $\bf i$ and  $\bf \frac{d(d+1)}{2}$ 
different resistors (coated with insulator) mutually connecting  all the 
nodes of the network (see Fig (2)).

As usual, total dissipative power {\bf TDP} in these networks can be written in 
terms of the  resistances and the currents flowing in them. But it is more 
convenient and also advantageous  throughout this article to express {\bf TDP} in 
terms of the inward flowing currents $\bf I_i,i=1,2,\cdots,d+1$.
In that case, it is clear that {\bf TDP} is a bilinear function  of the input
currents with the coefficients which have the dimensions of the resistance.
 
Hence we call these coefficients, connection resistances denoted 
by $R_{jk},\; j,k=1,2,\cdots, d+1$. 
Therefore, {\bf TDP} of the network assumes the following form
\begin{equation}
{\bf TDP}(network)=\sum_{j,k=1}^{(d+1)}R_{jk}I_jI_k.
\end{equation}
It is clear from equation (3-1) that $\bf R_{jk}$ is symmetric with respect to 
the interchange of indices $\bf i$ and $\bf j $.  Also the diagonal elements 
$\bf R_{jj},j=1,2,\cdots,d+1$ can be eliminated from the expression (3-1), if
 we use  Kirchhoff's current law  for the input currents

\begin{equation}
\sum_{j=1}^{(d+1)}I_j=0.
\end{equation}
Thus, The expression (3-1) takes the following form  

\begin{equation}
{\bf TDP}(network)=-\sum_{j\ne k=1}^{(d+1)}R_{jk}I_jI_k=-2\sum_{k>j=1}^{(d+1)}R_{jk}I_jI_k.
\end{equation}

From positive definiteness of {\bf TDP} for all arbitrary values of input inward
flowing currents consistent with Kichhoff's current law, it follows that 
all connection resistances are positive, that is we have:

$$
R_{jk}>0 \;\;\;\;\; for\; all\; k>j=1,2,\cdots,d+1.
$$
From the form of the {\bf TDP} given in (3-3), it also follows that there is a bijective map
between these sets of the independent connection resistances 
$\bf \{R_{jk}, k>j=1,2,\cdots,d+1\}$ and  $\bf \frac{d(d+1)}{2}$  mutual resistors of 
$\bf (d+1)$-simplex network.
Accordingly, these independent connection resistances can represent the mutual 
resistors of the network and in the case of an anisotropic network the 
connection resistances will be different .  Consequently, for the 
investigation of the restoration of macroscopic isotropy in $\bf (d+1)$-simplex 
fractal resistor lattices,  by real space renomalization group method 
, we need to know the recursion relations between the connection  resistances 
of a given generation and the connection  resistances of one generation below it. 
 
These recursion relations can easily be obtained if we compare  the total 
dissipative power {\bf TDP} of generation $\bf n$ given in (3-3) with the same quantity, 
calculated  as sum of power of its $\bf (n-1) $th generated  subfractals  which can be 
expressed as a function of connection resitances of 
generation $\bf n-1$, provided that in calculating the power of its subfractals, the inner
inward flowing currents are stated in terms of input currents. 
To determine these currents it is convenient to denote
 the $\bf j$-th inward flowing current of subfractal corresponding to
the partition $\bf \lambda_1,\lambda_2, \cdots, \lambda_{d+1}$ by $\bf I_{\lambda_1,
\lambda_2, \cdots ,\lambda_{d+1} (\lambda_1, \cdots ,\lambda_{j-1}, \lambda_{j}+1, \lambda_{j+1}, \cdots , \lambda_{d+1})}$.
Thence $\bf I_j$, the $\bf j$-th inward flowing current of $\bf (d+1)$-simplex 
fractal, is given by

\begin{eqnarray}
\nonumber
I_{0,0, \cdot\cdot,0, \underbrace{1}_{j-th} ,0, \cdot\cdot, 0}(0,0, \cdot\cdot,0, \underbrace{2}_{j-th},0, \cdot\cdot, 0)=I_j.
\end{eqnarray}

To determine the inner inward flowing currents,
 besides applying Kirchhoff's current law at each node and 
subfractal, we have to minimize the total dissipative power
 of $\bf (d+1)$-simplex fractal of generation $\bf n$, calculated as the sum of the {\bf TDP} of
its subfractals as: 

\begin{eqnarray}
\nonumber
\hspace{-15mm}\sum_{_{_{_{_{_{_{\hspace{16mm}sum\;over\hspace{1mm} partition \hspace{1mm}of\hspace{1mm} (b-1)}}}}}}}
\hspace{-22mm}\sum_{j,k=1}^{d+1}R_{jk}(n-1)I_{\lambda_1,\cdots,\lambda_{d+1}}
(\lambda_1,\cdot\cdot,\lambda_j+1,\cdot\cdot,\lambda_{d+1})
I_{\lambda_1,\cdot\cdot,\lambda_{d+1}}
(\lambda_1,\cdot\cdot,\lambda_k+1,\cdot\cdot,\lambda_{d+1}) \nonumber \\
\sum_{sum\;over\hspace{1mm} partition \hspace{1mm} of \hspace{1mm}(b-1)}
-2\mu_{\lambda_1,\cdot\cdot,\lambda_{d+1}}I_{\lambda_1,\cdot\cdot,\lambda_{d+1}}(\lambda_1,\cdot\cdot,\lambda_k+1,\cdot\cdot,\lambda_{d+1}) \nonumber \\
\sum_{sum\;over\hspace{1mm}partition\hspace{1mm}of\hspace{1mm}b} \nonumber
-2\nu_{\eta_1,\cdot\cdot,\eta_{d+1}}I_{\eta_,\cdot\cdot,\eta_{k-1},\cdot\cdot,\eta_{d+1}}(\eta_1,\cdot\cdot,\eta_{d+1}),
\\ \hspace{-10mm}
\end{eqnarray}
where $\bf \mu_{\lambda_1,...,\lambda_{d+1}}$ and $\bf \nu_{\eta_1,...,\eta_{d+1}}$ 
 are lagrange multipliers due to Kirchhoff's law on each subfractal, and also on
each node,respectively.
Minimizing the  expression (3-4), we get linear equations between inner input  
flowing currents and lagrange multipliers  
together with the Kirchhoff's law for each subfractal and each vertex, 
respectively.  Solving the equations thus obtained  we can write all inner 
inward flowing currents as a linear function in terms of input ones. 
Substituting the expressions thus obtained  for the inner currents in Eq. (3-4), 
we determine {\bf TDP} of generation $\bf n$ which is obviously a bilinear function of  
input currents with coefficients which are in general very involved 
functions of the connection resistances of the generation $\bf n-1$. Comparing the 
final result with the expression (3-4), connection resistance of generation $\bf n$ as its coefficient, 
we get  the required transformation between connection resistances of 
generations $\bf n$ and $\bf n-1$, respectively:

\begin{equation}
R_{jk}(r+1))=f_{jk}(R_{lm}(r)_{m>l}),\;\;\;\;\; k>j=1,2,\cdot\cdot,d+1.   
\end{equation}   
Here in this article, we show that the power and the anisotropy suppression 
exponents can be calculated, without having any knowledge of the explicit 
form of the functions $\bf f_{jk}$. All we need to know is some general 
properties of these functions which can be obtained rather easily from some 
physical requirements and also from dimensional analysis: these functions 
are homogeneous functions of degree one mapping positive connection 
resistances of generation 
$\bf n-1$ into positive connection resistances of generation 
$\bf n$, that is they form positive homogeneous map of degree one.
   
All connection resistances are positive; none of them can be negative or 
zero. The physical reason behind it is that if, for example, the connection
resistance $\bf R_{jk}$ becomes negative or if it vanishes, then for inward
flowing currents $\bf I_j=-I_k=I$, and $\bf I_l=0$ if $\bf l\neq j\neq k$, we obviously get
negative or zero power which is not physical in either cases. Analogously, we can 
rather easily deduce that none of them can be infinite, since all resistors 
of the network are finite, otherwise we will have infinite total dissipative 
power which is not again physical. 
 
 Definitely the transformation (3-5) is monotonically increasing, since by 
 increasing the connection resistances
  at a given generation $\bf n-1$, without  changing the input currents, 
 the total dissipative power of generation $\bf n$ will increase, that is  the
  connection resistances of generation $\bf n$ will increase.
Naturally, under the action of the point group $\bf S_{(d+1)}$,\cite{Ham} the connection
 resistances  simply permute among themselves. For example, the exchange
 of the vertices $\bf j$ and $\bf k$ in $\bf (d+1)$-simplex induces the following
 transformation among the connection resistances:

\begin{eqnarray}
\nonumber       
R_{jk} \longrightarrow & R_{kj}&=R_{jk} \nonumber \\
R_{jl} \longrightarrow & R_{lk}&\;\;\;\;\;\;\;\;\;\;\;\;\;\;\;\;\;\;\;\;\;\;\;\;  for\;\; l\neq j\neq k \nonumber \\
R_{kl} \longrightarrow & R_{lj}&\;\;\;\;\;\;\;\;\;\;\;\;\;\;\;\;\;\;\;\;\;\;\;\; for\;\; l\neq j\neq k \nonumber \\
R_{lm} \longrightarrow & R_{lm}&\;\;\;\;\;\;\;\;\;\;\;\;\;\;\;\;\;\;\;\;\;\;\;\; for\;\; l\neq m\neq j\neq k.
\end{eqnarray} 
As an example, we give the explicit form of the transformation for the special case of   
$\bf d=2$ and $\bf b=2$

\begin{eqnarray}
\nonumber
R_{12}(n)=\frac{R_{12}[R_{12}(n-1)+2R_{13}(n-1)+ 2R_{23}(n-1)]}{R_{12}(n-1)+R_{13}(n-1)+ R_{23}(n-1)}, \nonumber \\
R_{13}(n)=\frac{R_{13}[R_{13}(n-1)+2R_{12}(n-1)+ 2R_{23}(n-1)]}{R_{12}(n-1)+R_{13}(n-1)+ R_{23}(n-1)}, \nonumber \\
R_{23}(n)=\frac{R_{12}[R_{23}(n-1)+2R_{12}(n-1)+ 2R_{13}(n-1)]}{R_{12}(n-1)+R_{13}(n-1)+ R_{23}(n-1)}.\nonumber 
\end{eqnarray}

\section{ Fixed Point of Recurrence Equation of Connection\newline Resistances}
\setcounter{equation}{0}

In this section we present a rigorous proof that the renormalization
group transformation of the connection resistances has a unique fixed
direction in the space of connection resistances. That is, all of the flows stemming from
the finite physical region of connection resistance space converge to
infinity  at a direction which has the same angle with all coordinates.

For simplicity we denote the connection resitances of generation $\bf n-1$ $\bf  R_{jk}(n-1) $ $\bf  (k, j=1,2,\cdots, d+1) $, by $\bf  X_{\alpha}  $ 
$\bf ( \alpha=1,2,\cdots , \frac{d(d+1)}{2})$
and the connection resistances of generation $\bf n$
$\bf R_{jk}(n)$ $\bf ( k, j=1,2,\cdots, d+1)$, 
 by $\bf X_{\alpha}^{\prime}$ 
$\bf  (\alpha=1,2,\cdots , \frac{d(d+1)}{2})$,
respectively. Then the transformations (3-5) can be written as

\begin{equation}
X_{\alpha}^{\prime}=f_{\alpha}(X_{\beta})\;\;\;\; for \;\; \alpha=1,2,\cdots ,   
\frac{d(d+1)}{2}.
\end{equation}
Now we consider $\bf X_{\alpha}> 0 $ 
$\bf ( \alpha=1,2,\cdots ,\frac{d(d+1)}{2}) $ as 
coordinates of the interior points of a cone in $\bf \frac{d(d+1)}{2}$ dimensional 
Euclidean space denoted by $\bf {\large \breve {\cal C}}$. Denoting the cone itself by
$\bf  {\large {\cal C}} $, the transformation (4-1) can be considered as the map of this
cone into itself:
\begin{equation}
\nonumber       
{\large {\cal C}}\stackrel{F}{ \longrightarrow }{\large {\cal C}},
\end{equation}
where we have denoted the extension of the map (4-1) over the cone itself by $\bf F$. 

From the action of the permutation group $\bf S_{(d+1)}$ on the space of
connection resistances (the cone $\bf {\large {\cal C}}$) given in (3-6), it follows that
the transformation (4-1) is equivariant with respect to the action of $\bf S_{(d+1)}$,
that is we have the following commutative diagram \cite {Ish}:

\begin{eqnarray}
\nonumber       
&{\large {\cal C}}&\stackrel{g}{ \longrightarrow} {\large {\cal C}} \nonumber \\
F &\downarrow& \;\;\;\;\;\;\;\downarrow F\;\;\;\;\;\;\;\;\;\;\;\;\;\;\;\;\;for\;every\; g\in S_{(d+1)}, \nonumber \\
&{\large {\cal C}}&\stackrel{g}{ \longrightarrow }{\large {\cal C}}\nonumber
\end{eqnarray}
or
\begin{eqnarray}
g(F({\large {\cal X}}))=F(g( {\large {\cal X}}))\;\;\;\;\;\;\;\;\;\;for\;every\; g\in S_{(d+1)}\;\; and
\;\; {\large {\cal X}}\in {\large {\cal C}}.
\end{eqnarray}

A cone in Euclidean space
has the following properties\cite{Vas}:   

1. $\bf {\large {\cal C}}$+$\bf  {\large {\cal C}} \subset  {\large {\cal C}}$
   
2. $\bf \lambda {\large {\cal C}} \subset  {\large {\cal C}}$   for all $\bf  \lambda >0$

3.  $\bf  {\large {\cal C}} \bigcap  {\large {\cal -C}} ={0}   $.\\
We denote interior of this cone by $\bf  {\large \breve {\cal C}}$. One can define an 
"order relation" as follows:

$\bf  {\large {\cal X}} \geq  {\large {\cal Y}}$ if $\bf  {\large {\cal X-Y}}\in 
{\large {\cal C}}$
and
$\bf  {\large {\cal X}} >  {\large {\cal Y}}$ if $\bf  {\large {\cal X-Y}}\in 
{\large \breve {\cal C}}$. If one defines the numbers $\bf m({\large {\cal X}},{\large {\cal Y}})$ and
$\bf M({\large {\cal X}},{\large {\cal Y}})$ such that

\begin{eqnarray}
\nonumber 
m({\large {\cal X}},{\large {\cal Y}})=max_i \frac{x_i}{y_i}\\
M({\large {\cal X}},{\large {\cal Y}})=min_i \frac{x_i}{y_i},
\end{eqnarray}
then for any $\bf {\large{\cal X}},{\large {\cal Y}} \in {\large \breve {\cal C}}$   
the following relation holds   
\begin{eqnarray}
m({\large {\cal X}},{\large {\cal Y}}){\large {\cal Y}}\leq {\large {\cal X}}\leq
M({\large {\cal X}},{\large {\cal Y}}){\large {\cal Y}}.
\end{eqnarray}
The above relation and also all the other theorems of this section have been proved  
in reference \cite{Vas}.
Using the numbers defined in (4-4), one can define the Hilbert metric 
$\bf d({\large {\cal X}},{\large {\cal Y}})$ for any  
 $\bf {\large{\cal X}},{\large {\cal Y}} \in {\large \breve {\cal C}}$   as
\begin{eqnarray}
d({\large {\cal X}},{\large {\cal Y}})=
\log[\frac{M({\large {\cal X}},{\large {\cal Y}})}
{m({\large {\cal X}},{\large {\cal Y}})}]=
\log[ max_{i,j}\frac{x_iy_j}{y_ix_j}],
\end{eqnarray}
with the usual property of a pseudometric on  $\bf {\large \breve {\cal C}}$ and a metric
on $\bf {\large \breve {\cal C}}\bigcap S(0,1)$, where $\bf S(0,1)$ denotes the set of points
of sphere of radius one with the origin as its center. 
The metric space $\bf {\large \breve {\cal C}}\bigcap S(0,1)$ is complete under Hilbert
metric(4-6).
Also, it is straightforward
to see that the following assertions about this metric are valid : 

1. For any $\bf {\large {\cal X}},{\large {\cal Y}} \in {\large \breve {\cal C}}$ and 
$\bf  a,b\in R $ we have
$$
d(a{\large {\cal X}},b{\large {\cal X}} )=d({\large {\cal X}},{\large {\cal X}}). 
$$

2. $\bf d({\large {\cal X}},{\large {\cal X}})$ =0 if and only if 
$\bf {\large {\cal X}}=\lambda{\large {\cal X}} $.

3. For any $\bf {\large {\cal X}},{\large {\cal Y}}\in {\large \breve {\cal C}}$ 
the metric
$\bf d({\large {\cal X}},{\large {\cal X}}) $ is finite.
A given map of the cone into itself:

$$
F: {\large {\cal C}} \longrightarrow {\large {\cal C}}
$$
is called positive homogeneous of degree $\bf n$ and monotonically increasing if,
 for all $\bf {\large {\cal X}}\in {\large {\cal C}} $ and $\bf a>0$, we have
$$
F(a{\large {\cal X}})=a^nF({\large {\cal X}}),
$$
and 
$$
 {\large {\cal X}}\leq {\large {\cal Y}}\Longrightarrow  
F({\large {\cal X}})\leq F({\large {\cal Y}}).  
$$
According to the arguments given in section {\bf III}, the transformation (3-5) or
 (4-1) is monotonically increasing, positive and  homogeneous map of degree one.  
 
Using the relation (4-5) one can prove that for a positive homogeneous map
of degree one and monotonically increasing map like the transformation (4-1), the
following inequality holds:
\begin{equation}
m({\large {\cal X}},{\large {\cal Y}})T({\large {\cal X}}) \leq 
T({\large {\cal X}})\leq.  
M({\large {\cal X}}),({\large {\cal Y}})T({\large {\cal Y}}).  
\end{equation}
It is straightforward to get the following inequality from the inequality (4-7)
\begin{equation}
m({\large {\cal X}},{\large {\cal Y}}) \leq 
m({\large {\cal TX}},{\large {\cal TY}})\leq
M({\large {\cal X}},{\large {\cal Y}})\leq 
M({\large {\cal TX}},{\large {\cal TY}}).  
\end{equation}
From the above inequality and also from the definition of Hilbert metric (4-6),
it follows that for all $\bf {\large {\cal X}},{\large {\cal Y}} \in 
{\large {\cal C}}\bigcap S(0,1)$ we have

\begin{equation}
d({\large {\cal TX}},{\large {\cal TY}})\leq
d({\large {\cal X}},{\large {\cal Y}}).
\end{equation}
Therefore, the transformation (4-1) satisfies the Lipschitz condition and is of
contractive type. Thus, according to the Principle of Contraction Mapping 
Theorem, the contracting mapping (4-1) has a unique fixed point 
$\bf {\large {\cal X}}_0$ in the complete metric space 
$\bf ({\large {\cal C}}\bigcap S(0,1), d)$ (d is the Hilbert metric given in (4-4)
and
\begin{equation}
\lim_{n\rightarrow \infty}
\overbrace{F(F(\cdots F(F}^{n}
({\large {\cal X}}))\cdots))={\large {\cal X}}_0,
\;\;\;\;\;\;\;\;\;\;\;
for\;every\;{\large {\cal X}}\in {\large {\cal C}}.
\end{equation}
But, because of the equivariant property (4-3)of the transformation (4-1), any fixed point of
the point group $\bf S_{(d+1)}$(or the stability point of the point group \cite{Ish}) will
definitely be the fixed point of the transformation (4-1) acting on the space
$\bf ({\large {\cal C}}\bigcap S(0,1))$. Obviously, the direction $\bf X_1=X_2=\cdot=X_{\frac{d(d+1)}{2}}$ is the only fixed point of the
permutation group $\bf S_{(d+1)}$ acting on the space
$\bf ({\large {\cal C}}\bigcap S(0,1))$.
Hence, because of the uniqueness of the fixed point of the transformation (4-1)
on the space $\bf ({\large {\cal C}}\bigcap S(0,1))$, the direction
$\bf X_1=X_2=\cdot=X_{\frac{d(d+1)}{2}}$ is the only fixed direction of the connection
resistances renormalization group transformation. This direction corresponds
to the isotropic $\bf \bf (d+1)$-simplex,  which indicates that the macroscopic
conductivity  becomes isotropic on large scales.

\section{ Determination of Inner Inward Flowing Currents\newline of Subfractals in
Isotropic State}
\setcounter{equation}{0}
In order to determine the inner inward flowing currents in terms of the input
currents $\bf I_j$ $\bf (j=1,2,\cdot,d+1)$ in isotropic state, we have to minimize
the {\bf TDP} given in (3-4). But here in isotropic state all connection
resistances are the same, hence they can be put equal to one in (3-4),
simply by rescaling the lagrange multipliers of current conservations of
vertices and subfractals. Now, by minimizing {\bf TDP}, we get the
following equation for $\bf I$
\begin{equation}
I_{\lambda_1,\cdots,\lambda_{d+1}}(\lambda_1,\cdots,\lambda_j+1,\cdots,\lambda_{d+1})
-\mu_{\lambda_1,\cdots,\lambda_j,\lambda_{d+1}}-nu_{\lambda_1,\cdots,\lambda_j+1,\cdots,\lambda_{j+1}}
=0
\end{equation}
together with the Kirchhoff's law for each subfractal and each vertex, respectively
\renewcommand{\theequation}{\thesection-\arabic{equation}{a}}

\begin{equation}
\sum_{j=1}^{d+1}I_{\lambda_1,\cdot\cdot,\lambda_j,\cdot\cdot,\lambda_{d+1}}(\lambda_1,\cdot\cdot,\lambda_j+1,\cdot\cdot,\lambda_{d+1})=0.
\end{equation}
\setcounter{equation}{1}
\renewcommand{\theequation}{\thesection-\arabic{equation}{b}}
\begin{equation}
\sum_{j=1}^{d+1}I_{\eta_1,\cdot\cdot,\eta_j-1,\cdot\cdot,\eta_{d+1}}(\eta_1,\cdot\cdot,\eta_j,\cdot\cdot,\eta_{d+1})=0.
\end{equation}
We assume the following ansatz for the Lagrange multipliers:
\setcounter{equation}{2}
\renewcommand{\theequation}{\thesection-\arabic{equation}{a}}
\begin{equation}
\mu_{\lambda_1,\lambda_2,\cdot\cdot,\lambda_{d+1}}=\sum_{k=1}^{d+1}a_{\lambda_1,\lambda_2,\cdot\cdot,\lambda_{d+1}}(\lambda_k)I_k     
\end{equation}
\setcounter{equation}{2}
\renewcommand{\theequation}{\thesection-\arabic{equation}{b}}
\begin{equation}
\nu_{\eta_1,\eta_2,\cdot\cdot,\eta_{d+1}}\sum_{k=1}^{d+1}b_{\eta_1,\eta_2,\cdot\cdot,\eta_{d+1}}(\eta_k)I_k  
\end{equation}
with $a_{\lambda_1,\lambda_2,\cdot\cdot,\lambda_{d+1}}(0)$ and 
$b_{\eta_1,\eta_2,\cdot\cdot,\eta_{d+1}}(0)$ taken  to be zero. 

Using the ansatz (5-3a) and (5-3b) in equation (5-1), the inflowing currents can 
be given in terms of $\bf a$ and $\bf b$ respectively, that is
\setcounter{equation}{3}
\renewcommand{\theequation}{\thesection-\arabic{equation}}
\begin{equation}
I_{\lambda_1,\cdot\cdot,\lambda_{d+1}}(\eta_1,\cdot\cdot,\eta_{d+1})=\sum_{k=1}^{d+1}a_{\lambda_1,\lambda_2,\cdot\cdot,\lambda_{d+1}}(\lambda_k)I_k     
+\sum_{k=1}^{d+1}b_{\eta_1,\eta_2,\cdot\cdot,\eta_{d+1}}(\eta_k)I_k.  
\end{equation}
Due to the $ \bf S_{(d+1)} $ permutation symmetry of $\bf (d+1)$-simplex fractal, the  parameters
$\bf a_{\lambda_1,\lambda_2,\cdot\cdot,\lambda_{d+1}}(\lambda_k)$ and
 $\bf b_{\eta_1,\eta_2,\cdot\cdot,\eta_{d+1}}(\eta_k)$ depend only on the corresponding
partition $\bf  \{\lambda_1,\lambda_2,\cdot\cdot,\lambda_{d+1}\}$ and 
$\bf \{\eta_1,\eta_2,\cdot\cdot,\eta_{d+1}\}$, respectively. They do not change under the permutation of 
$\bf  \lambda_i$ or $\eta_i $ within a given partition. From now on, as far as $a$ and $b$ are 
concerned, only nonzero values are going to be quoted in their partition.

Actually one could write the currents in terms of input ones as in (5-4) 
by simply using the  symmetry of simplex fractal,  and the minimization of power  
is not required. 
Finally $\bf a$ and $\bf b$ can be determined through the equations (5-2a) and (5-2b). Obviously the number of equations are the same as the number of 
unknowns, hence the unknowns $\bf a$ and $\bf b$ can be determined uniquely. Here 
we determine the currents only for $\bf b=2,3,4$ and $\bf  5 $,  respectively. 

Let us first consider the case where $\bf b=2 $

$$
\hspace{-70mm}I_{0,0,\cdot\cdot,0,\underbrace{1}_{j-th},0,\cdot\cdot,0}(0,0,\cdot\cdot,0,\underbrace{2}_{j-th},0,\cdot\cdot,0)=I_j
$$
$$
\hspace{-10mm}I_{0,0,\cdot\cdot,0,\underbrace{1}_{j-th},0,\cdot\cdot,0}(0,0,\cdot\cdot,0,\underbrace{1}_{j-th},0,\cdot\cdot,0,\underbrace{1}_{k-th},0,\cdot\cdot,0)=a_1(1)I_J+b_1(1)I_j+b_(1)1I_k
$$
Using equation (5-2b) we have

$$
\hspace{-17mm}a_1(1)+2b_1(1)=0
$$
and from equation (5-2a) we get

$$
1+da_1(1)+(d-1)b_1(1)=0.
$$
Solving the above equations we get the following result  

$$
\hspace{-34mm}I_{0,\cdot\cdot,0,\underbrace{1}_{j-th},0,\cdot\cdot,0}(0,\cdot\cdot,0,\underbrace{1}_{j-th},0,\cdot\cdot,0,\underbrace{1}_{k-th},0,\cdot\cdot,0)=\frac{(I_k-I_j)}{(d+1)}.
$$

Via the the procedure explained above, we can similarly calculate the inner
inward flowing currents corresponding to ${\bf b=3,4}$ and ${\bf
b=5}$, where the details of calculation appear in Appendices I, II and
III, respectively and below we quote only the results:

I: Inner inward flowing currents corresponding to decimation number 
${\bf b=3}$

$$
\hspace{-52mm}I_{0,\cdot\cdot,0,\underbrace{2}_{j-th},0,\cdot\cdot,0}(0,\cdot\cdot,0,\underbrace{2}_{j-th},0,\cdot\cdot,0,\underbrace{1}_{k-th},0,\cdot\cdot,0)=
$$
$$
-\frac{2d+5}{(2d+3)(d+1)}I_j+\frac{3}{(2d+3)(d+1)}I_k
$$
$$
\hspace{-29mm}\hspace{-9mm}I_{0,\cdot\cdot,0,\underbrace{1}_{j-th},0,\cdot\cdot,0,\underbrace{1}_{k-th},0,\cdot\cdot,0}(0,\cdot\cdot,0,\underbrace{2}_{j-th},0,\cdot\cdot,0,\underbrace{1}_{k-th},0,\cdot\cdot,0)=
$$
$$
\frac{2d+5}{(2d+3)(d+1)}I_j -\frac{3}{(2d+3)(d+1)}I_k
$$
$$
\hspace{-29mm}\hspace{-9mm}I_{0,\cdot\cdot,0,\underbrace{1}_{j-th},0,\cdot\cdot,0,\underbrace{1}_{k-th},0,\cdot\cdot,0}(0,\cdot\cdot,0,\underbrace{1}_{j-th},0,\cdot\cdot,0,\underbrace{1}_{k-th},0,\cdot\cdot,0)=
$$
$$         
\frac{2}{(2d+3)(d+1)}(2I_l-I_j-I_k).
$$
\\
\\

II: Inner inward flowing currents corresponding to decimation number 
${\bf b=4}$

$$
\hspace{-65mm}I_{0,\cdot\cdot,0,\underbrace{3}_{j-th},0,\cdot\cdot,0}(0,\cdot\cdot,0,\underbrace{3}_{j-th},0,\cdot\cdot,0,\underbrace{1}_{k-th},0,\cdot\cdot,0)=
$$

$$
-\frac{8d^3+52d^2+5(25d+21)}{P}I_j+\frac{25d+49}{P}I_k
$$

$$
\hspace{-50mm}I_{0,\cdot\cdot,0,\underbrace{2}_{j-th},0,\cdot\cdot,0,\underbrace{1}_{k-th},0,\cdot\cdot,0}(0,\cdot\cdot,0,\underbrace{3}_{j-th},0,\cdot\cdot,0,\underbrace{1}_{k-th},0,\cdot\cdot,0)=
$$

$$
\frac{8d^3+52d^2+125d+105}{P}I_j-\frac{25d+49}{P}I_k
$$

$$
\hspace{-51mm}I_{0,\cdot\cdot,0,\underbrace{2}_{j-th},0,\cdot\cdot,0,\underbrace{1}_{k-th},0,\cdot\cdot,0}(0,\cdot\cdot,0,\underbrace{2}_{j-th},0,\cdot\cdot,0,\underbrace{2}_{k-th},0,\cdot\cdot,0)=
$$

$$
-\frac{12d^2+79d+91}{P}(I_j-I_k)
$$

$$
\hspace{-27mm}I_{0,\cdot\cdot,0,\underbrace{2}_{j-th},0,\cdot\cdot,0,\underbrace{1}_{k-th},0,\cdot\cdot,0}(0,\cdot\cdot,0,\underbrace{2}_{j-th},0,\cdot\cdot,0,\underbrace{1}_{k-th},0,\cdot\cdot,0,\underbrace{1}_{l-th},0,\cdot\cdot,0)=
$$

$$
-8\frac{d^2+6d+7}{P}I_j-4\frac{3d+7}{P}I_k+2\frac{19d+35}{P}I_l
$$

$$
\hspace{-13mm}I_{0,\cdot\cdot,0,\underbrace{1}_{j-th},0,\cdot\cdot,0,\underbrace{1}_{k-th},0,\cdot\cdot,0,\underbrace{1}_{l-th},0,\cdot\cdot,0}(0,\cdot\,0,\underbrace{2}_{j-th},0,\cdot\,0,\underbrace{1}_{k-th},0,\cdot\,0,\underbrace{1}_{l-th},0,\cdot\,0)=
$$

$$
16\frac{d^2+6d+7}{P}I_j-2\frac{13d+21}{P}(I_k+I_l)
$$

$$
\hspace{-2mm}I_{0,.,0,\underbrace{1}_{j-th},0,.,0,\underbrace{1}_{k-th},0,.,0,\underbrace{1}_{l-th},0,.,0}(0,.,0,\underbrace{1}_{j-th},0,.,0,\underbrace{1}_{k-th},0,.,0,\underbrace{1}_{l-th}
,0,\cdot,\underbrace{1}_{m-th},0,.,0)=
$$

$$
-4\frac{4d+7}{P}(I_j+I_k+I_l-3I_m)
$$

where $ P$ is defined as 

$$
P=(8d^3+44d^2+81d+49)(d+1).
$$
\\
\\

III: Inner inward flowing currents corresponding to decimation number 
${\bf b=5}$
$$
\hspace{-65mm}I_{0,.,0,\underbrace{4}_{j-th},0,\cdot\cdot,0}(0,\cdot\cdot,0,\underbrace{4}_{j-th},0,\cdot\cdot,0,\underbrace{1}_{k-th},0,\cdot\cdot,0)=
$$
$$
\frac{192d^6+2720d^5+16332d^4+53648d^3+102215d^2+106746d+47255}{Q}I_j
$$
$$
-3\frac{542d^3+3803d^2+8576d+6275}{Q}I_k
$$
$$
\hspace{-50mm}I_{0,\cdot\cdot,0,\underbrace{3}_{j-th},0,\cdot\cdot,0,\underbrace{1}_{k-th},0,\cdot\cdot,0}(0,\cdot\cdot,0,\underbrace{4}_{j-th},0,\cdot\cdot,0,\underbrace{1}_{k-th},0,\cdot\cdot,0)=
$$
$$
-\frac{192d^6+2720d^5+16332d^4+53648d^3+102215d^2+106746d+47255}{Q}
$$
$$
+3\frac{542d^3+3803d^2+8576d+6275}{Q}I_k
$$
$$
\hspace{-51mm}I_{\scriptsize{0,\cdot\cdot,0,\underbrace{3}_{j-th},0,\cdot\cdot,0,\underbrace{1}_{k-th},0,\cdot\cdot,0}}(\scriptsize{0,\cdot\cdot,0,\underbrace{3}_{j-th},0,\cdot\cdot,0,\underbrace{2}_{k-th},0,\cdot\cdot,0})=
$$
$$
\frac{288d^5+4104d^4+24466d^3+70139d^2+94822d+48213}{Q}I_j
$$
$$
-\frac{600d^4+8602d^3+36869d^2+62290d+36303}{Q}I_k
$$
$$
\hspace{-29mm}I_{\scriptsize{0,\cdot\cdot,0,\underbrace{3}_{j-th},0,\cdot\cdot,0,\underbrace{1}_{k-th},0,\cdot\cdot,0}}
(\scriptsize{0,\cdot\cdot,0,\underbrace{3}_{j-th},0,\cdot\cdot,0,\underbrace{1}_{k-th},0,\cdot\cdot,0,\underbrace{1}_{l-th},0,\cdot\cdot,0})=
$$
$$
2\frac{96d^5+1312d^4+7426d^3+20691d^2+27782d+14261}{Q}I_j
$$
$$
+2\frac{300d^3+2462d^2+6245d+5043}{Q}I_k
$$
$$
-2\frac{1326d^3+8947d^2+19483d+13782}{Q}I_l
$$
$$
\hspace{-50mm}I_{\scriptsize{0,\cdot\cdot,0,\underbrace{2}_{j-th},0,\cdot\cdot,0,\underbrace{2}_{k-th},0,\cdot\cdot,0}}
(\scriptsize{0,\cdot\cdot,0,\underbrace{3}_{j-th},0,\cdot\cdot,0,\underbrace{2}_{k-th},0,\cdot\cdot,0})=
$$
$$
-\frac{288d^5+4104d^4+24466d^3+70139d^2+94822d+48213}{Q}I_j
$$
$$
-\frac{600d^4+8602d^3+36869d^2+62290d+36303}{Q}I_k
$$
$$
\hspace{-35mm}I_{\scriptsize{0,\cdot\cdot,0,\underbrace{2}_{j-th},0,\cdot\cdot,0,\underbrace{2}_{k-th},0,\cdot\cdot,0}}
(\scriptsize{0,\cdot\cdot,0,\underbrace{2}_{j-th},0,\cdot\cdot,0,\underbrace{2}_{k-th},0,\cdot\cdot,0,\underbrace{1}_{l-th},0,\cdot\cdot,0})=
$$
$$
2\frac{144d^4+1896d^3+8260d^2+14607d+9059}{Q}I_j
$$
$$
+\frac{144d^4+1896d^3+8260d^2+14607d+9059}{Q}I_k
$$
$$
-2\frac{784d^3+5144d^2+10907d+7507}{Q}I_l
$$
$$
\hspace{-15mm}I_{0,\cdot\cdot,0,\underbrace{2}_{j-th},0,\cdot\cdot,0,\underbrace{1}_{k-th},0,\cdot\cdot,0,\underbrace{1}_{l-th},0,\cdot\cdot,0}
(\scriptsize{0,\cdot\,0,\underbrace{3}_{j-th},0,\cdot\,0,\underbrace{1}_{k-th},0,\cdot\,0,\underbrace{1}_{l-th},0,\cdot\,0})=
$$
$$
2\frac{1026d^3+6485d^2+13238d+8739}{Q}(I_k+I_l)
$$
$$
-4\frac{96d^5+1312d^4+7426d^3+20691d^2+27782d+14261}{Q}I_j
$$
$$
\hspace{-15mm}I_{\scriptsize{0,\cdot\cdot,0,\underbrace{2}_{j-th},0,\cdot\cdot,0,\underbrace{1}_{k-th},0,\cdot\cdot,0,\underbrace{1}_{l-th},0,\cdot\cdot,0}}
(\scriptsize{0,\cdot\,0,\underbrace{2}_{j-th},0,\cdot\,0,\underbrace{2}_{k-th},0,\cdot\,0,\underbrace{1}_{l-th},0,\cdot\,0})=
$$
$$
2\frac{312d^4+4084d^3+16326d^2+26083d+14489}{Q}I_j
$$
$$
-2\frac{228d^4+2990d^3+12293d^2+20345d+11774}{Q}I_k
$$
$$
+\frac{784d^3+5144d^2+10907d+7507}{Q}I_l
$$
$$
\hspace{-2mm}I_{\scriptsize{0,\cdot\cdot,0,\underbrace{2}_{j-th},0,\cdot\cdot,0,\underbrace{1}_{k-th},0,\cdot\cdot,0,\underbrace{1}_{l-th},0,\cdot\cdot,0}}
(\scriptsize{0,\cdot\cdot,0,\underbrace{2}_{j-th},0,\cdot\cdot,0,\underbrace{1}_{k-th},0,\cdot\cdot,0,\underbrace{1}_{l-th}
,0,\cdot,\underbrace{1}_{m-th},0,\cdot\cdot,0})=
$$
$$
8\frac{48d^4+596d^3+2389d^2+3899d+2238}{Q}I_j
$$
$$
+6\frac{152d^3+1062d^2+2361d+1691}{Q}I_k
$$
$$
+\frac{152d^3+1062d^2+2361d+1691}{Q}I_l
$$
$$
-2\frac{316d^3+2041d^2+4273d+2908}{Q}I_m
$$
$$
\hspace{-15mm}I_{\scriptsize{0,\cdot\cdot,0,\underbrace{1}_{j-th},0,\cdot\cdot,0,\underbrace{1}_{k-th},0,\cdot\cdot,0,\cdot\cdot,0,\underbrace{1}_{l-th},0,\cdot\cdot,0,\cdot\cdot,0,\underbrace{1}_{m-th},0,\cdot\cdot,0}}
(\scriptsize{0,\cdot\cdot,0,\underbrace{2}_{j-th},0,\cdot\cdot,0,\underbrace{1}_{k-th},0,\cdot\cdot,0,\underbrace{1}_{l-th},0,\cdot\cdot,0,\cdot\cdot,0,\underbrace{1}_{m-th},0,\cdot\cdot,0})=
$$
$$
-24\frac{48d^4+596d^3+2389d^2+3899d+2238}{Q}I_j
$$
$$
-\frac{164d^3+979d^2+1912d+1217}{Q}(I_k+I_l+I_m)
$$
$$
\hspace{-15mm}I_{\scriptsize{0,\cdot\cdot,0,\underbrace{1}_{j-th},0,\cdot\cdot,0,\underbrace{1}_{k-th},0,\cdot\cdot,0,\underbrace{1}_{l-th},0,\cdot\cdot,0,\cdot\cdot,0,\underbrace{1}_{m-th},0,\cdot\cdot,0}}
(\scriptsize{ 0,\cdot\cdot,0,\underbrace{1}_{j-th},0,\cdot\cdot,0,\underbrace{1}_{k-th},0,\cdot\cdot,0,\underbrace{1}_{l-th},0,\cdot\cdot,0,\underbrace{1}_{m-th},0,\cdot\cdot,0,\underbrace{1}_{n-th},0,\cdot\cdot,0})=
$$
$$
12\frac{96d^3+604d^2+1237d+825}{Q}(I_j+I_k+I_l+I_m-4I_n)
$$
where $ Q$ is defined as
$$
Q=192d^7+2720d^6+16332d^5+53648d^4+103841d^3+118155d^2+72983d+18825.
$$

\section{ Scaling Exponent of Anisotropy Suppression of\newline $\bf (d+1)$-Simplex Fractal Conductor Network}
\setcounter{equation}{0}

To investigate the abolition of anisotropy and also in order
to calculate the scaling exponent of its suppression, we linearize
the recursion map (4-1) near the fixed direction (isotropy state)
of this map:

$$
\lim_{n\rightarrow \infty}R_{jk}(n)=R\;\;\;\; for\;\; k>j=1,2,\cdot\cdot,d+1.
$$
This leads us to write
\begin{equation}
R_{jk}(n)=R+\varepsilon_{jk}(n);\;\;\; for\;\; k>j=1,2,\cdot\cdot,d+1
\end{equation}
with $\bf \varepsilon_{jk}(n)$ as an infinitesimal deviation of the connection
resistances of generation $\bf n$ from the isotropic state, for large values
of $\bf n$.
Now, all we need to know is the recursion relations between the deviation of
connection resistances of the generation $\bf n$
 and the infinitesimal deviation
of connection resistances of the generation $\bf n-1$,
for large values of $\bf n$.

These recursion relations can easily be obtained, if we compare the
deviation of {\bf TDP} from the isotropic state of generation $\bf n$ with the
deviation of the same quantity, calculated  as the sum of deviation of {\bf TDP}
of its subfractals of generation $\bf n-1$.
Clearly the deviation of {\bf TDP} of generation $\bf n$ can be obtained from
the expression (3-3), provided that in (3-3) we replace the connection
resistances with the deviation of the connection resistances of generation
$\bf n$. Also {\bf TDP} of generation $\bf n$ is the sum of {\bf TDP} of subfractals generations $\bf n-1$,
where again the latter can be obtained from (3-3), if we replace in (3-3)
the input currents with the inner inflowing currents (which have been
expressed in terms of input currents in section V) and the connection
resistances with the deviation of the generation $\bf n-1$, respectively.
Proceeding as above we obtain the recursion relations of the following form
for the deviation of connection resistances of generations $\bf n-1$ and $\bf n$ in a
$\bf (d+1)$-simplex fractal conductor network, for large values of $\bf n$:

\begin{eqnarray}
\nonumber
_{j \neq k}\varepsilon_{jk}(n)=f(d,b)_{j \neq k}\varepsilon_{jk}(n-1)
+g(d,b)(\sum_{l=1 \neq j \neq k}^{d+1}\varepsilon_{jl}(n-1)
+\sum_{l=1 \neq j \neq k}^{d+1}\varepsilon_{lk}(n-1)) \nonumber \\
+g(d,b)(\sum_{l\neq m=1 \neq j \neq k}^{d+1}\varepsilon_{lm}(n-1)).
\end{eqnarray}
For a given value of $\bf j \neq k $ we denote $\bf \varepsilon_{jk}(n)$($\varepsilon_{jk}(n-1)$) by $ X(X^{\prime}) $. Next we assume that 
$\bf \varepsilon_{jl}(n)$($\bf \varepsilon_{jl}(n-1)$ 
and $\bf \varepsilon_{lk}(n)$ $\bf (\varepsilon_{lk}(n-1))$ with ($\bf  l\neq j\neq k $)
are all equal which are denoted by $\bf Y$($\bf Y^{\prime}$). Finally we assume that
the remaining deviation of connection resistances, that is, 
$\bf \varepsilon_{lm}(n)$($\bf \varepsilon_{lm}(n-1)$) with ($\bf m\neq l\neq j\neq k$) 
are all equal which are denoted by $\bf Z(\bf Z^{\prime)}$. Then the recursion relations 
(6-2) take the following form:
\begin{eqnarray}
\nonumber
\hspace{-100mm}\left(\begin{array}{c}X^{\prime}\\Y^{\prime}\\Z^{\prime}\\ \end{array} \right ) = \nonumber\\
\nonumber
\vspace{3mm}
\hspace{-10mm}\left(\begin{array}{ccc}
\nonumber
f(d,b) & 2(d-1)g(d,b)& (d-2)(d-1)h(d,b) \nonumber \\
g(d,b) & f+g(d-1)+2h(d-2) & g(d-2)+h(d-3)(d-2) \nonumber\\                                               
2h(d,b) & 4g(d,b)+4(d-3)h(d,b) & f(d,b)+2(d-3)g(d,b)+(d-4)(d-3)h(d,b)\end{array} \right )\left(\begin{array}{c}X\\Y\\Z \end{array} \right).\nonumber 
\nonumber\\
\hspace{155mm}(6-3)\nonumber
\end{eqnarray}
The following eigen-values are obtained by diagonalizing the $\bf 3\times3$ matrix (6-3)
. The eigen-values are quoted in decreasing order as:
\begin{eqnarray}
\nonumber
\lambda_{max}&=&\frac{d^2h(d,b)+d(4g(d,b)-3h(d,b))+2f(d,b)-4g(d,b)+2h(d,b)}{2} \nonumber\\
\lambda_{midle}&=&d(g(d,b)-h(d,b))+f(d,b)-3g(d,b)+2h(d,b)  \nonumber\\
\lambda_{min}&=&f(d,b)-2g(d,b)+h(d,b),
\end{eqnarray}
where the corresponding eigen-vectors are given as the rows of the following 
matrix: 
\begin{eqnarray}
\vspace{-30mm}
\left ( \begin{array}{ccc} 1 & 1 & 1
\\ 2(d-1)  & d-3 & -2  \\ (d-1)(d-2) &-(d-2)&2 \end{array} \right ).  
\end{eqnarray}
We see that the eigen-directions do not depend on the decimation number $\bf b$
but rather only on the dimension $\bf d$. This is again due to equivariancy of
the map (4-1) with respect to the action of the point group $\bf S_{(d+1)}$ on the space
 of the connection resistances given in  (4-3).

As expected, the maximum eigen-value corresponds to isotropy state,
which gives the power scaling exponent of $\bf (d+1)$-simplex fractal 
conductor networks \cite{Ste,Roux,Pak,Jaff}. Therefore, anisotropy vanishes with a scaling 
exponent which can be obtained in terms of the eigenvalues in the 
 following way: from the recurence relation (3-5) and its linearized form (6-2), it follows that, 
for large values of $\bf n$ we have
\begin{equation}
\lim_{n\longrightarrow \infty}R_{jk}\sim=L_n^{D_2},\;\;\;\;\;\ for\;every\;j\neq k=1,2,\cdot,d+1
\end{equation}
where $\bf L_n=b^n$, and the power scaling exponent $\bf D_2$ is defined as: 
\begin{equation} 
D_2(d,b)=\frac{\log\lambda_{max}}{\log b}
\end{equation}
and
\begin{eqnarray}
\lim_{n\longrightarrow \infty}\frac{R_{jk}}{R_{jl}}-1\sim=
L_n^{\bar{\lambda}}\;\;\;\;\;\; for\;every\;j\neq k\neq l=1,2,\cdot,d+1\nonumber\\
\lim_{n\longrightarrow \infty}\frac{R_{jk}}{R_{lm}}-1\sim=L_n^{\bar{\lambda}}\;\;\;\;\;\; for\;every\;j\neq k\neq l\neq m=1,2,\cdot,d+1\nonumber\\
\end{eqnarray}
where the scaling exponent of suppression of the anisotropy, $\bf  \bar{\lambda}$ 
, is  defined as                             
\begin{equation} 
\bar{\lambda}(d,b)=
\frac{\log
\frac{maximum\;eigenvalue}{the\; next \;greatest\; eigenvalue}}
{\log b}=
\frac{\log\frac{\lambda_{max}}{\lambda_{midle}}}{\log b}.
\end{equation}
In the remaining part of this section  we quote the results for $\bf b=2,3,4$ and
$\bf 5$, respectively.
$\bf I:b=2$
\begin{eqnarray}
\nonumber
f(d,2)=\frac{5d+3}{(d+1)^2}\nonumber\\
g(d,2)=\frac{d-1}{(d+1)^2}\nonumber\\
h(d,2)=\frac{-2}{(d+1)^2}\nonumber\\
\nonumber\\
D_2(d,2)=\frac{\log{\frac{d+3}{d+1}}}{\log{2}}\nonumber\\
\nonumber\\
\bar{\lambda}(d,2)=\frac{\log{\frac{d+3}{d+2}}}{\log{2}}\nonumber
\\
\end{eqnarray}

$\bf II:b=3 $
\begin{eqnarray}
\nonumber
f(d,3)&=&\frac{(8d^3+88d^2+145d+59)}{2d+3)^2(d+1)^2}\nonumber\\
g(d,3)&=&\frac{4d^3+20d^2+d-24}{(2d+3)^2(d+1)^2}\nonumber\\
h(d,3)&=&\frac{-(4d^2+24d+25)}{(2d+3)^2(d+1)^2}\nonumber\\
\nonumber\\
D_2(d,3)&=&\frac{\log\frac{2d^2+9d+19}{(2d+3)(d+1)}}{\log 3}\nonumber\\
\nonumber\\
\bar{\lambda}(d,3)&=&\frac{log\frac{(2d+3)(2d^2+9d+19)}{4d^3+20d^2+41d+31}}{\log3}\nonumber
\\
\end{eqnarray}
\vspace{10mm}
{\bf III:b=4}
\begin{eqnarray}
\nonumber
f(d,4)&=&\frac{128d^7+1792d^6+13936d^5+59116d^4+137757d^3+175421d^2+113267d+28567}
{(8d^4+52d^3+125d^2+130d+49)^2}\nonumber\\
g(d,4)&=&\frac{64d^7+832d^6+4768d^5+13448d^4+16313d^3-701d^2-18501d-11319}
{(8d^4+52d^3+125d^2+130d+49)^2}\nonumber\\
h(d,4)&=&\frac{-(64d^6+896d^5+5664d^4+19096d^3+35061d^2+32970d+12397)}
{(8d^4+52d^3+125d^2+130d+49)^2}\nonumber\\
\nonumber\\
D_2(d,4)&=&\frac{\log\frac{8d^4+68d^3+253d^2+588d+539}{8d^4+52d^3+125d^2+130d+49}}{\log 4}\nonumber\\
\nonumber\\
\bar{\lambda}(d,4)&=&\frac{\log\frac{8d^4+68d^3+253d^2+588d+539}{8d^4+52d^3+125d^2+130d+49}}{\log 4}\nonumber
\\
\end{eqnarray}
\vspace{10mm}
{\bf IV:b=5}
\begin{eqnarray}
\nonumber
\hspace{-30mm}f(d,5)=\frac{1}{Q^2}(73728d^{13}+2162688d^{12}+29576192d^{11}+258155264d^{10}+1614743456d^9 \nonumber\\
\vspace{15mm} +7530179904d^8+26333589428d^7+68567523880d^6+131200269465d^5 \nonumber\\
\vspace{15mm} +180815964435d^4+173716650934d^3+109891587638d^2+40940417277d \nonumber\\
\vspace{15mm}+6768087791) \nonumber\\
\nonumber\\   
\hspace{-30mm}g(d,5)=\frac{1}{Q^2}(36864d^{13}+1044480d^{12}+13706752d^{11}+110574336d^{10}+607189008d^9\nonumber\\
\vspace{15mm}+2357625920d^8+6492213656d^7+12314821608d^6+14686625629d^5\nonumber\\
\vspace{15mm}+7431447086d^4-6360742466d^3-13852397352d^2-9571181547d\nonumber\\
-2499415382)\nonumber\\
\nonumber\\
\vspace{-30mm}h(d,5)=\frac{-1}{Q^2}(36864d^{12}+1081344d^{11}+14788096d^{10}+125353216d^9+732200464d^8\nonumber\\
\vspace{15mm}+3083843024d^7+9521524696d^6+21544010108d^5+35237710633d^4\nonumber\\
\vspace{15mm}+40459001364d^3+30870831766d^2+14032614408d+2871299265)\nonumber\\
\nonumber\\
\vspace{-35mm}D_2(d,5)=\frac{\
log{\frac{192d^7+3104d^6+22348d^5+95720d^4+280525d^3+559419d^2+652155d+320017}
{Q}}}{\log{5}}\nonumber\\
\nonumber\\
\vspace{-35mm}\bar{\lambda}(d,5)=\frac{ \log{\frac{(192d^7+3104d^6+22348d^5
+95720d^4+280525d^3+559419d^2+652155d+320017)Q}
{(d+1)P}}}{\log{5}}\nonumber
\\
\end{eqnarray}
with $\bf P$ and $\bf Q$ defined as
\begin{eqnarray}
\vspace{-30mm}P=(36864d^{13}+1044480d^{12}+13706752d^{11}+110574336d^{10}+614600976d^9\nonumber\\
\vspace{30mm}+2499189440d^8+7689210552d^7+18190236812d^6+33121369305d^5\nonumber\\
\vspace{30mm}+45749851193d^4+46378189714d^3+32473949342d^2+13985550557d\nonumber\\
+2781136877)\nonumber
\end{eqnarray}
and
$$
Q=(192d^6+2528d^5+13804d^4+39844d^3+63997d^2+54158d+18825)(d+1).
$$

It is straightforward to see that these results will also hold true  for
$\bf (d+1)$-honeycomb fractal conductor network with decimation number $\bf b$,
which can be constructed from a given $\bf (d+1)$-simplex fractal conductor
network, simply by replacing the resistors in the links with the resistors
which connect the center of a subfractal to its vertices, (see Fig. {\bf 2})
where, this has also been shown for $\bf d=2$ and $\bf b=2$ case in reference \cite{Pak}.
\\
\\
\\
\\

Appendix I: Calculation of currents of ${\bf b=3}$.

Here in this Appendix we give the detail of calculation of inner inward
flowing currents corresponding to decimation number ${\bf b=3}$

Following the procedure of section IV, for ${\bf b=3}$ we have

$$
\hspace{-78mm}I_{0,\cdot\cdot,0,\underbrace{2}_{j-th},0,\cdot\cdot,0}(0,%
\cdot\cdot,0,\underbrace{3}_{j-th},0,\cdot\cdot,0)=I_j
$$
$$
\hspace{-64mm}I_{0,\cdot\cdot,0,\underbrace{2}_{j-th},0,\cdot\cdot,0}(0,%
\cdot\cdot,0,\underbrace{2}_{j-th},0,\cdot\cdot,0,\underbrace{1}%
_{k-th},0,\cdot\cdot,0)=
$$
$$
\hspace{-39mm} a_2(2)I_j+b_{21}(2)I_j+b_{21}(1)I_k
$$
$$
\hspace{-49mm}I_{0,\cdot\cdot,0,\underbrace{1}_{j-th},0,\cdot\cdot,0,
\underbrace{1}_{k-th},0,\cdot\cdot,0}(0,\cdot\cdot,0,\underbrace{2}%
_{j-th},0,\cdot\cdot,0,\underbrace{1}_{k-th},0,\cdot\cdot,0)=
$$
$$
\hspace{-30mm}a_{11}(1)(I_j+I_k)+b_{21}(2)I_j+b_{21}(1)I_k
$$
$$
\hspace{-29mm}I_{0,\cdot\cdot,0,\underbrace{1}_{j-th},0,\cdot\cdot,0,
\underbrace{1}_{k-th},0,\cdot\cdot,0}(0,\cdot\cdot,0,\underbrace{1}%
_{j-th},0,\cdot\cdot,0,\underbrace{1}_{k-th},0,\cdot\cdot,0,\underbrace{1}%
_{l-th},0,\cdot\cdot,0)=
$$
$$
\hspace{-29mm}a_{11}(1)(I_j+I_k)+b_{111}(1)(I_j+I_k+I_l).
$$
Using equation (5-2a) in subfractal ${\bf (0,\cdot\cdot,0,\underbrace{2}%
_{j-th},0,\cdot\cdot,0)}$, we get

$$
\hspace{-52mm}1+d(a_2(2)+b_{21}(2))-b_{21}(1)=0,
$$
also using equation (5-2a) in subfractal ${\bf %
(0,...,0,1_j,0,...,1_k,0,...,0)}$ we get
$$
\hspace{-29mm}(d+1)a_{11}(1)+b_{21}(1)+b_{21}(2)+(d-2)b_{111}(1)=0,
$$
also, for vertices equation (5-2b) gives

\begin{eqnarray*}
& &a_2(2)+2b_{21}(2)+a_{11}(1)=0\\
& &a_{11}(1)+2b_{21}(1)=0\\
& &2a_{11}(1)+3b_{111}(1)=0.
\end{eqnarray*}

By solving the above equations we can determine inner inward flowing currens
corresponding to decimation number ${\bf b=3}$ which is given in section V.
\\
\\
\\

Appendix II: Calculation of currents of ${\bf b=4}$.

Here in this Appendix we give the detail of calculation of inner inward
flowing currents corresponding to decimation number ${\bf b=4}$

Similarly, following the procedure of section IV, for ${\bf b=4}$ we have

$$
\hspace{-81mm}I_{0,\cdot\cdot,0,\underbrace{3}_{j-th},0,\cdot\cdot,0}(0,%
\cdot\cdot,0,\underbrace{4}_{j-th},0,\cdot\cdot,0)=I_j
$$
$$
\hspace{-65mm}I_{0,\cdot\cdot,0,\underbrace{3}_{j-th},0,\cdot\cdot,0}(0,%
\cdot\cdot,0,\underbrace{3}_{j-th},0,\cdot\cdot,0,\underbrace{1}%
_{k-th},0,\cdot\cdot,0)=
$$
$$
\hspace{-81mm} a_3(3)I_j+b_{31}(3)I_j+b_{31}(1)I_k
$$
$$
\hspace{-50mm}I_{0,\cdot\cdot,0,\underbrace{2}_{j-th},0,\cdot\cdot,0,
\underbrace{1}_{k-th},0,\cdot\cdot,0}(0,\cdot\cdot,0,\underbrace{3}%
_{j-th},0,\cdot\cdot,0,\underbrace{1}_{k-th},0,\cdot\cdot,0)=
$$
$$
\hspace{-64mm}a_{21}(2)I_j+a_{21}(1)I_k+b_{31}(3)I_j+b_{31}(1)I_k
$$
$$
\hspace{-51mm}I_{0,\cdot\cdot,0,\underbrace{2}_{j-th},0,\cdot\cdot,0,
\underbrace{1}_{k-th},0,\cdot\cdot,0}(0,\cdot\cdot,0,\underbrace{2}%
_{j-th},0,\cdot\cdot,0,\underbrace{2}_{k-th},0,\cdot\cdot,0)=
$$
$$
\hspace{-69mm}a_{21}(2)I_j+a_{21}(1)I_k+b_{22}(2)(I_j+I_k)
$$
$$
\hspace{-29mm}I_{0,\cdot\cdot,0,\underbrace{2}_{j-th},0,\cdot\cdot,0,
\underbrace{1}_{k-th},0,\cdot\cdot,0}(0,\cdot\cdot,0,\underbrace{2}%
_{j-th},0,\cdot\cdot,0,\underbrace{1}_{k-th},0,\cdot\cdot,0,\underbrace{1}%
_{l-th},0,\cdot\cdot,0)=
$$
$$
\hspace{-54mm}a_{21}(2)I_j+a_{21}(1)+b_{211}(2)I_j+b_{211}(1)(I_k+I_l)
$$
$$
\hspace{-15mm}I_{0,\cdot\cdot,0,\underbrace{1}_{j-th},0,\cdot\cdot,0,
\underbrace{1}_{k-th},0,\cdot\cdot,0,\underbrace{1}_{l-th},0,\cdot%
\cdot,0}(0,\cdot\,0,\underbrace{2}_{j-th},0,\cdot\,0,\underbrace{1}%
_{k-th},0,\cdot\,0,\underbrace{1}_{l-th},0,\cdot\,0)=
$$
$$
\hspace{-47mm}a_{111}(1)(I_j+I_k+I_l)+b_{211}(2)I_j+b_{211}(1)(I_k+I_l)
$$
$$
\hspace{-2mm}I_{0,.,0,\underbrace{1}_{j-th},0,.,0,\underbrace{1}%
_{k-th},0,.,0,\underbrace{1}_{l-th},0,.,0} (0,.,0,\underbrace{1}%
_{j-th},0,.,0,\underbrace{1}_{k-th},0,.,0,\underbrace{1}_{l-th},0,\cdot,
\underbrace{1}_{m-th},0,.,0)=
$$
$$
\hspace{-44mm}a_{111}(1)(I_j+I_k+I_l)+b_{1111}(1)(I_j+I_k+I_l+I_m).
$$
Now, imposing Kirchhoff's law on subfractals and vertices, we get the
following equations for ${\bf a}$ and ${\bf b}$

\begin{eqnarray*}
& &1+da_3(3)+b_{31}(3)-b_{31}(1)=0\\
& &(d+1)a_{21}(2)+b_{31}(3)+b_{22}(2)+(d-1)b_{211}(2)-b_{211}(1)=0\\
& &(d+1)a_{21}(2)+b_{31}(1)+b_{22}(2)+(d-2)b_{211}(1)=0\\
& &(d+1)a_{111}(2)+b_{211}(2)+(d-3)b_{1111}(1)=0\\
& &a_{21}(1)+2b_{31}(1)=0  \\
& &a_{3}(3)+b_{21}(2)+2b_{31}(3)=0\\
& &a_{21}(2)+a_{21}(1)+2b_{22}(2)=0   \\
& &2a_{21}(2)+a_{111}(1)+3b_{211}(2)=0\\
& &a_{21}(1)+a_{111}(1)+3b_{211}(1)=0  \\
& &3a_{111}(1)+4b_{1111}(1)=0.
\end {eqnarray*}

By solving the above equations we can determine inner inward
flowing currents corresponding to decimation number ${\bf b=4}$
which appear in section V.
\\
\\
\\

Appendix III: Calculation of currents of ${\bf b=5}$.

Here in this Appendix we give the detail of calculation of inner inward
flowing currents corresponding to decimation number ${\bf b=5}$

Finally following the procedure of section IV, for ${\bf b=5}$ we have

$$
\hspace{-81mm}I_{0,\cdot\cdot,0,\underbrace{4}_{j-th},0,\cdot\cdot,0}(0,%
\cdot\cdot,0,\underbrace{5}_{j-th},0,\cdot\cdot,0)=I_j
$$
$$
\hspace{-65mm}I_{0,\cdot\cdot,0,\underbrace{4}_{j-th},0,\cdot\cdot,0}(0,%
\cdot\cdot,0,\underbrace{4}_{j-th},0,\cdot\cdot,0,\underbrace{1}%
_{k-th},0,\cdot\cdot,0)=
$$
$$
\hspace{-81mm} a_4(4)I_j+b_{41}(4)I_j+b_{41}(1)I_k
$$
$$
\hspace{-50mm}I_{0,\cdot\cdot,0,\underbrace{3}_{j-th},0,\cdot\cdot,0,
\underbrace{1}_{k-th},0,\cdot\cdot,0}(0,\cdot\cdot,0,\underbrace{4}%
_{j-th},0,\cdot\cdot,0,\underbrace{1}_{k-th},0,\cdot\cdot,0)=
$$
$$
\hspace{-64mm}a_{31}(3)I_j+a_{31}(1)I_k+b_{41}(4)I_j+b_{41}(1)I_k
$$
$$
\hspace{-51mm}I_{0,\cdot\cdot,0,\underbrace{3}_{j-th},0,\cdot\cdot,0,
\underbrace{1}_{k-th},0,\cdot\cdot,0}(0,\cdot\cdot,0,\underbrace{3}%
_{j-th},0,\cdot\cdot,0,\underbrace{2}_{k-th},0,\cdot\cdot,0)=
$$
$$
\hspace{-69mm}a_{31}(3)I_j+a_{31}(1)I_k+b_{32}(3)I_j+b_{32}(3)I_k
$$
$$
\hspace{-24mm}I_{0,\cdot\cdot,0,\underbrace{3}_{j-th},0,\cdot\cdot,0,
\underbrace{1}_{k-th},0,\cdot\cdot,0}(0,\cdot\cdot,0,\underbrace{3}%
_{j-th},0,\cdot\cdot,0,\underbrace{1}_{k-th},0,\cdot\cdot,0,\underbrace{1}%
_{l-th},0,\cdot\cdot,0)=
$$
$$
\hspace{-54mm}a_{31}(3)I_j+a_{31}(1)+b_{311}(3)I_j+b_{311}(1)(I_k+I_l)
$$
$$
\hspace{-45mm}I_{0,\cdot\cdot,0,\underbrace{2}_{j-th},0,\cdot\cdot,0,
\underbrace{2}_{k-th},0,\cdot\cdot,0}(0,\cdot\cdot,0,\underbrace{3}%
_{j-th},0,\cdot\cdot,0,\underbrace{2}_{k-th},0,\cdot\cdot,0)=
$$
$$
\hspace{-64mm}a_{22}(2)(I_j+I_k)+b_{32}(3)I_j+b_{32}(2)I_k
$$
$$
\hspace{-25mm}I_{0,.,0,\underbrace{2}_{j-th},0,.,0,\underbrace{2}%
_{k-th},0,.,0}(0,.,0,\underbrace{2}_{j-th},0,.,0,\underbrace{2}_{k-th},0,.,0,
\underbrace{1}_{l-th},0,.,0)=
$$
$$
\hspace{-64mm}a_{22}(2)(I_j+I_k)+b_{221}(2)(I_j+I_k)+b_{221}(1)I_l
$$
$$
\hspace{-15mm}I_{0,.,0,\underbrace{2}_{j-th},0,.,0,\underbrace{1}%
_{k-th},0,.,0,\underbrace{1}_{l-th},0,.,0}(0,.,0,\underbrace{3}_{j-th},0,.,0,
\underbrace{1}_{k-th},0,.,0,\underbrace{1}_{l-th},0,.,0)=
$$
$$
\hspace{-47mm}%
a_{211}(2)I_j+a_{211}(1)(I_k+I_l)+b_{311}(3)I_j+b_{311}(1)(I_k+I_l)
$$
$$
\hspace{-15mm}I_{0,.,0,\underbrace{2}_{j-th},0,.,0,\underbrace{1}%
_{k-th},0,.,0,\underbrace{1}_{l-th},0,.,0}(0,.,0,\underbrace{2}_{j-th},0,.,0,
\underbrace{2}_{k-th},0,.,0,\underbrace{1}_{l-th},0,.,0)=
$$
$$
\hspace{-47mm}%
a_{211}(2)I_j+a_{211}(1)(I_k+I_l)+b_{221}(2)(I_j+I_k)+b_{221}(1)I_l
$$
$$
\hspace{-2mm}I_{0,.,0,\underbrace{2}_{j-th},0,.,0,\underbrace{1}%
_{k-th},0,.,0,\underbrace{1}_{l-th},0,.,0} ({0,.,0,\underbrace{2}%
_{j-th},0,.,0,\underbrace{1}_{k-th},0,.,0,\underbrace{1}_{l-th},0,\cdot,
\underbrace{1}_{m-th},0,.,0})=
$$
$$
\hspace{-44mm}%
a_{211}(2)I_j+a_{211}(1)(I_k+I_l)+b_{2111}(2)I_j+b_{2111}(1)(I_k+I_l+I_m)
$$

$$
\hspace{-15mm}I_{0,.,0,\underbrace{1}_{j-th},0,.,0,\underbrace{1}%
_{k-th},0,.,0,\underbrace{1}_{l-th},0,.,0,.,0,\underbrace{1}_{m-th},0,.,0} ({%
0,.,0,\underbrace{2}_{j-th},0,.,0,\underbrace{1}_{k-th},0,.,0,\underbrace{1}%
_{l-th},0,.,0,.,0,\underbrace{1}_{m-th},0,.,0})=
$$
$$
\hspace{-47mm}%
a_{111}(1)(I_j+I_k+I_l+I_m)+b_{2111}(2)I_j+b_{2111}(1)(I_k+I_l+I_m)
$$
$$
\hspace{-15mm}I_{0,.,0,\underbrace{1}_{j-th},0,.,0,\underbrace{1}%
_{k-th},0,.,0,\underbrace{1}_{l-th},0,.,0,.,0,\underbrace{1}_{m-th},0,.,0} ({%
0,.,0,\underbrace{1}_{j-th},0,.,0,\underbrace{1}_{k-th},0,.,0,\underbrace{1}%
_{l-th},0,.,0,.,0,\underbrace{1}_{m-th},0,.,0\underbrace{1}_{n-th},0,.,0})=
$$
$$
\hspace{-47mm}a_{111}(1)(I_j+I_k+I_l+I_m)+b_{11111}(1)(I_j+I_k+I_l+I_m+I_n).
$$
Again imposing Kirchhoff's law on subfractals and vertices, we get the
following equations for ${\bf a}$ and ${\bf b}$ \newpage
\begin {eqnarray*}
& &1+da_4(4)+b_{41}(3)-b_{41}(1)=0 \\
& &(d+1)a_{31}(3)+b_{41}(4)+b_{32}(3)+(d-1)b_{311}(3)-b_{311}(1)=0\\
& &(d+1)a_{31}(1)+b_{41}(1)+b_{32}(2)+(d-2)b_{311}(1)=0\\
& &(d+1)a_{22}(1)+b_{32}(3)+b_{32}(2)+(d-1)b_{221}(2)-b_{221}(1)=0\\
& &(d+1)a_{211}(2)+b_{311}(3)+2b_{221}(2)+(d-2)b_{2111}(2)-b_{2111}(1)=0\\
& &(d+1)a_{211}(2)+b_{311}(1)+b_{221}(2)+b_{221}(1)+(d-3)b_{2111}(1)=0\\
& &(d+1)a_{1111}(2)+b_{2111}(2)+3b_{2111}(1)+(d-4)b_{11111}(1)=0\\
& &a_{4}(4)+a_{31}(3)+2b_{41}(4)=0  \\
& &a_{31}(1)+2b_{41}(1)=0\\
& &a_{31}(3)+a_{22}(2)+2b_{32}(3)=0  \\
& &a_{31}(1)+a_{22}(2)+2b_{32}(2)=0\\
& &2a_{31}(3)+a_{211}(2)+3b_{311}(3)=0  \\
& &a_{31}(1)+a_{211}(1)+3b_{311}(1)=0\\
& &a_{22}(2)+a_{211}(2)+a_{211}(1)+3b_{221}(2)=0   \\
& &2a_{211}(1)+3b_{221}(1)=0\\
& &3a_{211}(2)+a_{1111}(1)+4b_{2111}(2)=0  \\
& &2a_{211}(1)+a_{1111}(1)+4b_{2111}(1)=0\\
& &4a_{1111}(1)+5b_{11111}(1)=0.
\end{eqnarray*}

By solving the above equations we can determine inner inward flowing currens
corresponding to decimation number ${\bf b=5}$ which appear in section V.

.
\vspace{10mm}

{\large \bf Conclusion}\\
Here in this work it has been rigorously shown that 
the macroscopic isotropy will be restored if the corresponding renormalization 
map between two different scales has properties such as: positivity, homogeneity
of first order and most of all monotonically increasing property. Obviously, homogeneity
is enough and the order of homogeneity does not play a very important role. It 
is clear that this can be true in many physical phenomena, where we quote only 
 very few of them here: Diffusion in inhomogeneous media \cite{Smith,Haus}, elasticity property
of rubber or the network of polymer chains\cite{Bas}, conductivity in random resitor network \cite{Cle}
, flux distribution in josephson junction networks\cite{Grim}. It would be rather interesting
to see whether there exist the restoration isotropy which is not due to
positive, homogeneous, and monotonically increasing renormalization map of two different scales.

\vspace{2cm}

{\bf {\large  ACKNOWLEDGEMENT}}

We wish to thank  Dr. S. K. A. Seyed Yagoobi for his careful reading the 
article and for his constructive comments.

\end{document}